\documentclass[conference,a4paper]{IEEEtran}
\makeatletter
\def\ps@headings{%
\def\@oddhead{\mbox{}\scriptsize\rightmark \hfil \thepage}%
\def\@evenhead{\scriptsize\thepage \hfil \leftmark\mbox{}}%
\def\@oddfoot{}%
\def\@evenfoot{}}
\makeatother
\usepackage{cite}
\usepackage[utf8]{inputenc}
\usepackage{graphicx}
\usepackage{float}
\usepackage{amsfonts}    % For the math symbols
\usepackage{caption}
\usepackage{subcaption}
\usepackage{algorithmic}    % For the pseudocode
\usepackage{algorithm}    % For the pseudocode (is actually forbidden by IEEEtran)
\usepackage{fixltx2e}       % For the 2 column figure
\begin{document}
\title{60 GHz MAC Standardization: Progress and Way Forward}
\author{\IEEEauthorblockN{Kishor Chandra\IEEEauthorrefmark{1}, Arjan Doff\IEEEauthorrefmark{1}, Zizheng Cao\IEEEauthorrefmark{2}, R. Venkatesha Prasad\IEEEauthorrefmark{1}, Ignas Niemegeers\IEEEauthorrefmark{1}}\\ 
\IEEEauthorblockA{\IEEEauthorrefmark{1}Faculty of Electrical Engineering, Mathematics and Computer Science, TU Delft, The Netherlands\\
Email: {k.chandra, i.g.m.m.niemegeers}@tudelft.nl, a.w.doff@student.tudelft.nl, rvprasad@ieee.org.}\\
\IEEEauthorblockA{\IEEEauthorrefmark{2}COBRA Institute, Eindhoven University of Technology, The Netherlands, Email: Z.Cao@tue.nl\\
}}
\maketitle
\begin{abstract}
Communication at mmWave frequencies has been the focus in the recent years. In this paper, we discuss standardization efforts in 60\,GHz short range communication and the progress therein. We compare the available standards in terms of network architecture, medium access control mechanisms, physical layer techniques and several other features. Comparative analysis indicates that IEEE 802.11ad is likely to lead the short-range indoor communication at 60\,GHz. We bring to the fore resolved and unresolved issues pertaining to robust WLAN connectivity at 60\,GHz. Further, we discuss the role of mmWave bands in 5G communication scenarios and highlight the further efforts required in terms of research and standardization.
\end{abstract}
\section{introduction}
Research interests in mmWave are being revived after more than 100 years when the first demonstration was done by Jagadish Chandra Bose at Royal Society. WiFi (IEEE 802.11) operating at 2.4/5 GHz has emerged as the most popular choice for wireless local area networks (WLANs) services in indoor environments and hotspots. In recent years, there have been several efforts to increase the data rate of WiFi. Many efforts such as using higher order modulation scheme (e.g., 64/256 QAM), multiple input multiple output (MIMO) and channel bonding techniques at the physical layer,  frame aggregation and,  service differentiation techniques at the MAC layer have been introduced to enhance the capacity of WiFi networks. As a result, data rates from 54 Mb/s (IEEE 802.11g) to 1 Gb/s (IEEE 802.11ac) have been successfully achieved.  On the other hand, rapid increase in the number of wireless mobile devices and new applications such as online gaming and uncompressed HD video streaming have led to exploding growth of Internet traffic. This unprecedented traffic growth requires multi Gb/s Wireless connectivity in the indoor environment, and WLANs at 2.4\,GHz and 5 GHz are not able to provide such a high data rate wireless connectivity due to scarcity of spectrum resources.

\begin{figure}[]
\centering
\includegraphics[width=0.4\textwidth]{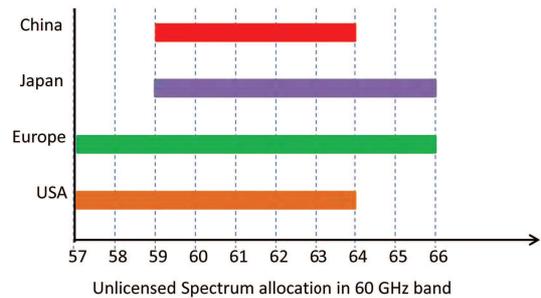}
\caption{Frequency allocation in different countries.}
\label{fig:60GHzbandsCountry}
\vspace{-7mm}
\end{figure}

Owing to this, there is a considerable interest in 60\,GHz band due to the availability of huge unlicensed spectrum band of around 5\,GHz (see Figure~\ref{fig:60GHzbandsCountry}). It has drawn much attention from industry, academia and standardization bodies. Because of availability of very high bandwidth, it has emerged as a potential choice for massive broadband wireless connectivity. This has led to several standardization efforts such as ECMA-387~\cite{ecma}, IEEE 802.15.3c~\cite{iee:IEEE802.15.3c}, WiGig and IEEE 802.11ad~\cite{ieee.ad2012}. ECMA-387 and IEEE 802.15.3c have not received much attention in terms of commercialization. However, IEEE 802.11ad is gaining the momentum recently in terms of product realization due to its backward compatibility with popular IEEE 802.11x protocols. Wilocity and Qualcomm have already demonstrated IEEE 802.11ad based chipsets. It has been anticipated that by 2017, IEEE 802.11ad will be an integral part of many consumer electronic devices and gadgets such as personal computers, tablets and mobile phones. 

Based on the fact that all of these standards aim to provide high speed short range communication at 60\,GHz for WLANs/WPANs, we study and compare different techniques and mechanisms proposed by them. These standards differ from each other in terms of medium access control (MAC), PHY techniques, network architectures and targeted applications. Owing to the special characteristics of 60\,GHz signals, there are many issues for deployment of 802.11ad WLANs. There are numerous articles available in the literature on the progress of IEEE 802.11 standards operating at 2.4 and 5\,GHz, but there are very few articles~\cite{shankar2012,compare11ad153c,baykas2011ieee} that report the progress in standardization activities at 60\,GHz frequency bands.
%For example, ECMA-387 uses a completely different PHY called STBC (Single Carrier Block Transmission); IEEE 802.15.3c has a centralized architecture and IEEE 802.11ad uses fallback option to 2.4\,GHz channel in the absence of sufficient 60\,GHz link budget.  

In addition, mmWave communication is seen as a potential candidate for 5G mobile communication systems. Apart from 60\,GHz frequency band, 28-32\,GHz, 38-42\,GHz bands are among the possible alternatives for high data rate 5G mobile/cellular networks. Though 5G networks employing mmWave communications will have relatively different requirements, but due to similarity in the spectral properties of signals in the mmWave bands -- existing 60\,GHz standards such IEEE 802.11ad and IEEE 802.15.3c can be used as the starting point to carve out the shape of future mmWave based 5G wireless communications.

In this article, we report the progress in 60\,GHz standardization and provide a comparative study of these standards highlighting their differences and similarities. Further, we discuss the challenges to be addressed for practical deployment of 60\,GHz Multi Gb/s WLAN  for seamless and  reliable connectivity. We also discuss the recent research activities and proposed solutions to address these challenges e.g., PHY and MAC layer enhancements proposed for uninterrupted and reliable WLAN services at 60\,GHz. Further we extend our discussions to the requirement and challenges in future mmWave based mobile networks.

The rest of the article is as follows. Sections~\ref{sec:ecma}, ~\ref{sec:15.3c} and ~\ref{sec:11ad} describe the ECMA-387, IEEE 802.15.3c and IEEE 802.11ad specifications, respectively. Section\ref{sec:compare_standards} provides the comparison of three standards in terms of various provisions. Section~\ref{sec:challengesWLAN} discusses the important issues for deployment of 60 GHz WLAN systems due to special characteristics of 60 GHz mmWave signals. Further, in Section~\ref{role_5G} we discuss about the need for integration of mmWave bands with lower frequency bands in 5G networks. Finally we conclude in Section~\ref{conc}.
\section{ECMA-387 Specifications}\label{sec:ecma}
ECMA-387 published by ETSI, defines 60\,GHz WPANs operating over four channels with a separation of $2.16$\,GHz within the frequency bands between $57.24$\,GHz -- $65.880$\,GHz. ECMA-387 specifies two types of devices, \textit{viz}, device Type A and device Type B. Type A device is expected to support  high data rates (up to $6.350$\,Gb/s), multi-level QoS, robust multipath performance and adaptive antenna arrays capable of beamforming and beamsteering. Type B device aims to be simple, low power, low cost, and targeted to be suitable for handheld devices supporting data rate up to $3.175$\,Gb/s. Channel bonding of adjacent channels is facilitated to increase the data rates of both the device types. The Type A PHY includes two general transmission schemes, namely Single Carrier Block Transmission (SCBT), also known as Single Carrier with Cyclic Prefix, and Orthogonal Frequency Division Multiplexing (OFDM).  The advantage of SCBT mode over OFDM is that it lowers the peak to average power ratio (PAPR) and hence preferred. On the other hand, Type B minimizes the complexity and power consumption of the receiver and may not support antenna training  for beamforming.
%\begin{figure}[]
%\centering
%\includegraphics[width=0.4\textwidth]{images/OOB.pdf}
%\caption{OOB control channel used by ECMA-387.}
%\label{fig:OOB}
%\vspace{-5mm}
%\end{figure}

ECMA-387 provides a decentralized MAC protocol for both the device types enabling coexistence, interoperability, QoS provisions and spatial reuse. This standard supports NoAck, ImmAck and BlockAck policies. Primarily, ECMA-387 standard supports a completely decentralized operation where each station sends its beacon over the discovery channel. There are two kind of devices: (i)~who can send beacons, and (ii)~who cannot send beacons. Coordination among beacon capable devices is fully distributed while in case of beacon sending and non-beacon devices existing together, a beacon sending device works as a controller. It also define protocol adaptation layer (PAL) which interacts with MAC layer through multiplexing sublayer to support different applications. To ensure communication among heterogeneous devices, a separate discovery channel is reserved. 
%ECMA-387 provides two power management modes in which a device can operate: active and hibernation. Devices in active mode transmit and receive beacons in every superframe. Devices in hibernation mode hibernate for multiple superframes and do not transmit or receive in those superframes. In addition, this standard provides facilities to support devices that sleep for portions of each superframe in order to save power.

To provide a better WPAN experience, ECMA-387 provides a low rate 2.4\,GHz control channel called out of band (OOB) control channel to support the unstable 60\,GHz channel. A MAC convergence layer  is defined to coordinate between 2.4\,GHz and 60\,GHz channels, and to support device discovery, synchronization, association control, service discovery, 60 GHz channel reservation and scheduling. There are two OOB operation modes \textit{viz}, ad hoc and infrastructure mode. In ad hoc mode there is no controller and each device sends OOB beacons periodically, while in infrastructure mode controller periodically sends the OOB beacons to which devices respond with association requests over 2.4\,GHz OOB channel and form the network.
 
%An important function of OOB control channel is to report the loss of 60\,GHz link to transmitter if adequate signal power is not received over 60\,GHz data link. Transmitter and receiver then switch over to 60\,GHz discover channel and restart the beamforming procedure to relinquish the 60\,GHz data link. Devices can also use OOB control channel to transmit acknowledgment frames for 60\,GHz data. Apart from the use of OOB control channel for WPAN management, ECMA-387 has laid provisions for using intermediate devices as amplify and forward relays if LOS connection is not feasible between a source and destination device due to blockage or bad channel conditions. In this case source device needs to identify a relay device through which it can reach the destination device, and it has to reserve the timeslots on agreed channel pairs between source to relay and relay to destination device. Thus using relay nodes, alternative paths can be found in case of non-availability of direct path between source and destination devices.
\section{IEEE 802.15.3c Specifications}\label{sec:15.3c} 
IEEE 802.15.3c was the first standard proposed by IEEE for 60\,GHz WPAN services. IEEE 802.15.3c defines three millimeter wave (mmWave) based PHY named as single carrier (SC) PHY, high-speed interface (HSI) PHY and audio visual (AV) PHY respectively. SC PHY mode, also known as office desktop model is designed to support low cost, low complexity while maintaining relatively high data rate to support high performance applications with data rate in excess of 3\,Gb/s and 5\,Gb/s respectively. HSI PHY is designed for devices with low latency, bidirectional high speed data and uses OFDM which is suitable for conference ad hoc user model with base rate for data at 1.54\,Gb/s and highest up to 5.77\,Gb/s. AV PHY is designed for typical audio video consumer electronics usage model. For these applications two different sub PHY modes are defined: high data rate PHY (HRP)  for video transmission and low data rate PHY (LRP) for control signal. Both modes use OFDM. Data rate for LRP is 2.5\,Mb/s to 10.2\,Mb/s and for HRP it is 0.952\,Gb/s to 3.807\,Gb/s. Common mode signaling (CMS) with data rate of 25.2\,Mb/s is supported by all the three PHYs for control and management frame transmissions.

The operating area of IEEE 802.15.3c is typically around a radius of 10\,m. This standard proposes a completely centralized network architecture where one device assumes the role of piconet coordinator (PNC) of the piconet. The piconet either operates in omni mode or in quasi-omni mode in which directional communication is supported.

IEEE 802.15.3c employs a hybrid MAC protocol which uses both the contention based access and fixed time division multiple access (TDMA) based medium access mechanisms.  Timing in IEEE 802.15.3c is based on the superframe (SF).  The superframe consists of three parts: beacon, contention access period (CAP ) and channel time allocation period (CTAP ). Beacon is used to communicate the timing allocations and management information for the piconet. CAP is used to communicate commands and asynchronous data if it is present in the superframe. Channel access mechanism used in CAP period is CSMA/CA. CTAP is used for isochronous data transmission. Channel time allocation in CTAP is purely TDMA which is allotted during CAP period. It is guaranteed that no other DEVs will compete for the channel during the indicated time duration of the CTA allotted to a DEV. In order to ensure reliable frame transmissions, IEEE 802.15.3c provides three acknowledgment mechanisms, namely, ImmAck, BlockACk and DelayACK. It also supports NoACK mode when acknowledgment is not sought after frame transmission.
\begin{figure}[]
\centering
\includegraphics[width=0.3\textwidth]{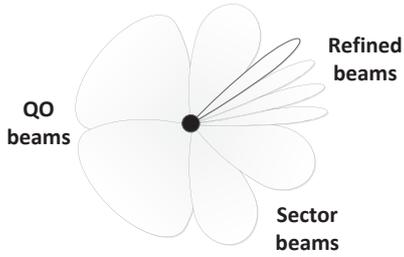}
\caption{Different beam levels in IEEE 802.15.3c and IEEE 802.11ad.}
\label{fig:beamforming_all_sectors}
\vspace{-7mm}
\end{figure}
\begin{table*}[]
%% increase table row spacing, adjust to taste
%\renewcommand{\arraystretch}{1.0}
% if using array.sty, it might be a good idea to tweak the value of
% \extrarowheight as needed to properly center the text within the cells
\caption{Comparison of PHY parameters.}
\label{table_parameters60GHz}
\centering
%% Some packages, such as MDW tools, offer better commands for making tables
%% than the plain LaTeX2e  
\begin{tabular}{|c||c||c||c|}
\hline
Parameter & ECMA-387 & IEEE 802.15.3c & IEEE 802.11ad  \\
\hline
Frequency band (GHz) &57-66&57-66&57-66\\
\hline
Channel bandwidth (GHz) & 2.160 &2.160 &2.160\\
\hline
Channel bonding & 2,3 or 4 channels can be aggregated & not allowed & not allowed\\
\hline
Control PHY rates (\,Mb/s) &397  &25.10 & 27.50\\
\hline
Highest data rate (\,Gb/s) &6.350  &5.70 &6.70\\
\hline
\end{tabular}
\end{table*}
To support directional communication, IEEE 802.15.3c provides a three tier beamforming mechanism (see Fig.~\ref{fig:beamforming_all_sectors}). Widest beamwidth level is called Quasi-omni (QO). Each QO level can have several sectors having narrow beamwidths. Further, each sector can be divided into very fine beams and called beam level. During beacon and CAP, QO level beamwidth is used for broadcasting management information and channel contention by devices, respectively. During CTAP periods, device pairs can further narrow down their beamwidths up to sector levels or high resolution beam levels. Beamforming mechanism is used to select the best transmit receive beam pairs at each level. Further, during data transmission, devices use special training packets to track best beam pairs in order to maintain the link quality. Since special training packets are used, it is called \textit{out packet training}.
\section{{IEEE 802.11{\textnormal{ad}} Specifications}} \label{sec:11ad}
\begin{figure}[b]
\centering
\includegraphics[width=0.4\textwidth]{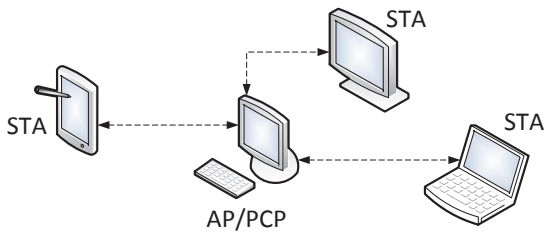}
\caption{An example of a IEEE 802.11ad architecture.}
\label{fig:WLAN_architecture}
\end{figure}

%\subsection{Network architecture}
% Before the 60 GHz band was introduced to \textsc{wlan} networks as an amendment to the IEEE 802.11 protocol, it was first used in an amendment to the IEEE 802.15 protocol in \textsc{wpan} networks.
The IEEE 802.11ad amendment requires the STAs to communicate independently of each other, therefore it uses a personal basic service set (PBSS). To assign basic timing to the STAs, one STA is required to be the PBSS central point (PCP), as shown in Figure~\ref{fig:WLAN_architecture}.

\begin{figure}[H]
\centering
\includegraphics[width=0.4\textwidth]{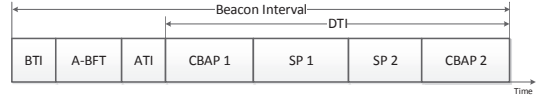}
\caption{A superframe of the MAC layer of the IEEE 802.11ad protocol.}
\label{fig:MACsuperframe_ad}
\vspace{-8mm}
\end{figure}

%\subsection{PHY layer}
The 802.11ad protocol has defined 3 different PHY structures: Control PHY, single carrier (SC) (with low-power SC) PHY and OFDM PHY \cite{ieee.ad2012}.
Control PHY operates at the lowest data rate, but uses the highest coding gain, such that it can be used for low signal-to-interference-plus-noise ratio (SINR) situations. It is used before a beamformed link is setup or for control frame transmissions.
SC PHY is designed for low power and low complexity transceivers. The last is OFDM PHY which can achieve the highest data rate.
%\begin{figure}[]
%\centering
%\includegraphics[width=0.4\textwidth]{images/PHYpacket_ad.pdf}
%\caption{The structure of a IEEE 802.11ad packet.}
%\label{fig:PHYpacket_ad}
%\end{figure}
%
% A packet in the PHY layer has a structure as shown in Figure~\ref{fig:PHYpacket_ad}. The first two fields are the short training field (STF) and channel estimation (CE) fields. They help with signal acquisition, automatic gain control (AGC) training, predicting the characteristics of the channel, frequency offset estimation and synchronization \cite{shankar2012}. The header contains general information about the packet, such as the modulations and coding schems (MCS), the size of the packet and also if the optional training fields are appended. The data field contains the MAC header and MAC data.
%\subsection{MAC layer}
The access methods used in 802.11ad comprises of both CSMA/CA and TDMA \cite{ieee.ad2012}.
A frame is referred to as a beacon interval (BI).
The structure of such a BI is shown in Figure~\ref{fig:MACsuperframe_ad}.
The BI consists of multiple parts. The first part is the beacon transmission interval (BTI), in which the PCP/AP transmits one or more beacons in different directions.
STAs willing to join the PBSS, can be trained in the association beamforming training (A-BFT) stage of the BI.
During announcement time (AT) the PCP/AP can transmit information to the STAs in a request/response fashion.
The main data transmission part is the data transmission interval (DTI), in which two periods are present.
The contention based access period (CBAP) and service period (SP) allows any frame exchange, including data transmissions \cite{cordeiro2010}.
Where CSMA/CA is used in CBAPs and TDMA is used in SPs.
It is possible to use any combination in the number and order of SPs and CBAPs in the DTI \cite{ieee.ad2012}. IEEE 802.11ad also provides dynamic channel allocation in which PCP/AP polls STAs either during CBAP or SP periods and grants channel access. During CBAP, EDCA mechanism can be used by an STA for prioritized channel access.

One of the main advantages of IEEE 802.11ad with respect to other protocols is that it has the capability to switch between 2.4/5 bands and 60\,GHz band transmissions.
This is called fast session transfer (FST), and it allows seamless connectivity.
This is a major cornerstone for 802.11ad since the link quality can quickly degrade in a 60\,GHz network due to movement or blockage.
FST can operate in both transparent and non-transparent mode.
The MAC address is the same in both bands if the \textsc{sta}s are in transparent operation and different in non-transparent operation.
FST also supports both simultaneous and non-simultaneous operation.
However, frequent switching from 60\,GHz band to 2.4/5\,GHz band can be annoying for users.
%An important role for the MAC layer is the seamless connection it should offer.In order to achieve this, FST needs to perform efficiently.
% DMG CTS
% DMG DTS
% DMG CF
%
%As part of the beam refinement protocol (\textsc{brp}) and beam tracking (\textsc{bt}) phase the IEEE 802.11ad protocol also offers the option to add additional training packets at the end of a packet. Using these training packets during beamforming training and tracking can also be referred to as in-packet training.
%The \textsc{agc} fields are added in order to account for the variety of the signal strengths when transmitting and receiving beam training fields.
%\subsection{Beamforming}
In order for devices to communicate at a high data rate the 802.11ad protocol employs beamforming. The beamforming setup consists of three phases similar to IEEE 802.15.3c. The first phase is the sector level sweep (SLS). Its purpose is to allow communication between two STAs. SLS is followed by beam refinement phase (BRP) in which STAs narrow down there beams.The different level beams can be seen in Figure~\ref{fig:beamforming_all_sectors}. The last phase is the beam tracking (BT) phase and it is done to further track the beams/channel.
\section{Comparative analysis}\label{sec:compare_standards}
\subsubsection*{\textbf{PHY layer}}
From the PHY layer perspective, all the three standards use several modulation and coding schemes some of which are similar and some differ with each other. ECMA-387 defines different PHYs for Type A and Type B devices with several combinations of modulation coding schemes. Apart from OFDM, Type A devices use single carrier block transmission (SCBT) scheme which is a unique feature of it and provide robust performance~\cite{ecma_evaluation}. It uses Reed solomon codes concatenated with convolution codes for device Type A. On the other hand, Type B devices use single carrier transmission with simple RS codes. Type A devices can use unequal error protection (UEP) while Type B devices cannot use it. Instead of device types, IEEE 802.15.3c defines three PHY modes (SC, AV and HSI) with several modulations and coding schemes (MCSs) suitable for wide range of applications.  Similar to IEEE 802.15.3c, IEEE 802.11ad defines three PHY modes named as Control PHY, SC PHY and OFDM PHY. It also proposes several MCSs with combination of different modulation formats and coding schemes. Table~\ref{table_parameters60GHz} summarizes the PHY parameters related to channelization and datarates of these three standards.
\begin{table*}[!t]
%% increase table row spacing, adjust to taste
%\renewcommand{\arraystretch}{1.0}
% if using array.sty, it might be a good idea to tweak the value of
% \extrarowheight as needed to properly center the text within the cells
\caption{Comparison on the basis of various mechanisms.}
\label{table_parameters60GHz_2}
\centering
%% Some packages, such as MDW tools, offer better commands for making tables
%% than the plain LaTeX2e 
\begin{tabular}{|p{3cm}|p{3cm}|p{3cm}|p{3cm}|} 
%\begin{tabular}{|c||c||c||c|}
\hline
Options & ECMA-387 & IEEE 802.15.3c & IEEE 802.11ad  \\
\hline
Network Architecture & Distributed & Centralized & Centralized\\
\hline
Medium access & CSMA/CA and TDMA & CSMA/CA and TDMA & CSMA/CA, TDMA, Polling \\
\hline
Dynamic Channel Access & No & No & Yes, PCP/AP can dynamically poll STAs during CBAP\\
\hline
Prioritized Medium Access & No & No & Yes, it uses EDCA mechanism proposed by IEEE 802.11e \\
\hline
Backward Compatibility & No & No & Yes, back compatible to IEEE 802.11 b/g/n/ac\\
\hline
Relay& yes & No & Yes\\
\hline
Fallback to 2.4\,GHz &No& No &Yes, Fast session transfer mechanism if 60\,Ghz link is not available\\
\hline
WPAN Management&Provision of 2.4\,GHz control plane& PNC operating over 60\,GHz&PCP/AP operating over 60\,GHz\\
\hline
\end{tabular}
\end{table*}
\subsubsection*{\textbf{Network Architecture}}
Primarily, ECMA-387 supports a completely distributed architecture without any controller (if only Type A devices are present). If Type B devices are also present, then one of the Type A device acts as a coordinator and network operates on master slave basis. On the other hand, IEEE 802.15.3c proposes completely centralized network architecture in which PNC coordinates communications among device pairs. Similarly, IEEE 802.11ad PBSS is centrally coordinated by PCP/AP. However, peer to peer communication is supported  by both IEEE 802.15.3c and IEEE 802.11ad.
\subsubsection*{\textbf{Device Discovery}}
In ECMA-387, Device discovery is achieved using beacon and polling frames. For same type of devices, device discovery is done using beacon frames employing CSMA/CA protocol. On the other hand, heterogeneous device discovery is done on a master slave basis using a polling protocol where Type A device works as a master and Type B device as a slave. In IEEE 802.15.3c and IEEE 802.11ad, PNC and PCP/AP periodically sends beacon frames in different QO directions. Once the STAs detect beacons, association requests are sent during association CAP and A-BFT periods by IEEE 802.15.3c and IEEE 802.11ad STAs using CSMA/CA protocol, respectively.
\subsubsection*{\textbf{MAC layer}}
ECMA-387 provides a distributed MAC mechanism in which following the device discovery phase device pairs reserve the channel for data transmission without intervention of any coordinator. On the other hand, IEEE 802.15.3c provides hybrid channel access mechanism in which devices use CSMA/CA or TDMA based channel access in CAP and CTAP durations, respectively. To reserve the TDMA slots, it uses CSMA/CA during CAP periods. Thus CAP period is used for data transmissions as well as for CTAP reservations.  IEEE 802.11ad also provides a hybrid channel access similar to IEEE 802.15.3c. CSMA/CA based data transmission is done during CBAP periods but the reservation of TDMA slots (called SPs) is done using polling by PCP/AP during ATI  period. Further, IEEE 802.11ad also has a provision for dynamic channel access --in which PCP/AP can dynamically poll STAs during CBAP or SP durations for fast channel access.
\subsubsection*{\textbf{Beamforming}}
For antenna training and tracking, ECMA-387 used special frames called TRN frames to determine the appropriate antenna weight vectors. Open loop and closed loop training and tracking mechanisms are given. In closed loop training, transmit antenna derives its weight vectors based on the feedback provided by receiver antenna while in open loop training there is no provision of feedback and same training weights are used of transmission and reception.
IEEE 802.15.3c provides a three level antenna training mechanism using beam codebooks~\cite{beamcodebookharada}, namely: (i)~best QO pattern training; (ii)~best sector level training; and (iii)~best beam pair training. During this procedure, it also uses special training frames. IEEE 802.11ad also uses a similar three level beamforming mechanism however it does not use special frames rather data frames are used and hence called \textit{in-packet training}. On the other hand, ECMA-387 and IEEE 802.15.3c training mechanism is called \textit{out packet training}. 
\subsubsection*{\textbf{Relay and Fallback option}}
Since 60\,GHz links are highly susceptible to link blockage due to channel variations -because of obstacles or misalignment of antenna beams, it is desirable to have alternate means to reclaim the lost links between devices. Also, device discovery and association becomes difficult due to directional communication at 60\,GHz. ECMA-387 has an option of using 2.4\,GHz OOB signaling for WPAN management. OOB is used for device discovery and association. Apart from OOB signaling it also proposes use of intermediate devices as relay if a 60\,GHz LOS link is broken to discover the alternate 60\,GHz path. IEEE 802.11ad also provides support for relays at 60\,GHz. Apart from relay support, it also provides fast session transfer mechanism to switch over 2.4 or 5\,GHz channel. On the contrary, IEEE 802.15.3c does not mention either support for relay or fallback option to lower frequency to relinquish the lost/blocked links.
\section{Way forward for 60\,GHz communication}
In this section, we discuss about the challenges in realization of WLAN connectivity at 60\,GHz frequency band followed by role of 60\,GHz technology in 5G wireless networks.
\subsection{Further challenges for robust WLAN connectivity at 60\,GHz}\label{sec:challengesWLAN}
In the preceding section we compared various aspects of three standards proposed for short range multi-Gb/s communications at 60\,GHz frequency bands. Various schemes proposed by these standards were discussed (see the Table~\ref{table_parameters60GHz_2}). However, to realize robust multi-Gb/s WLAN connectivity at 60\,GHz frequency bands like that of 2.4\,GHz bands is still a challenge. The main issues are: (i)~severe blockage of signals due to obstacles and (ii)~link outage due to mobility while using directional antennas.  IEEE 802.11ad has already emerged as the most favored 60\,GHz standard among device manufacturers. Hence it is desirable to further strengthen IEEE 802.11ad so that a reliable and robust WLAN service similar to WiFi can be delivered at 60\,GHz frequency bands. To tackle the link blockage, there is a provision for relay devices so that the alternate path can be used. However, further research is required in this domain. A cooperative MAC protocol using intermediate STAs as relay nodes is proposed in~\cite{DCTMAC}. The cooperating relaying enhances the performance of IEEE 802.11ad and extends the communication range. A method for optimum beamwidth selection for IEEE 802.11ad PCP/APs was presented in~\cite{chandra2014adaptive}. It was shown that how beamwidth selection can improve the performance of IEEE 802.11ad networks employing CSMA/CA MAC protocol. Further Kim, et al., have proposed a scheme for relay selection in IEEE 802.11ad multihop network while maximizing the video quality~\cite{relayAFMolish}. The relay selection depends on the video quality achieved in a multihop IEEE 802.11ad network. Relays play important role if LOS connection is blocked or if the source and destination STAs are far apart. For seamless user experience, intelligent relay selection mechanisms are required so that smooth link transition can be facilitated to users without any interruption in service delivery. Spatial diversity is used to combat the human induced shadowing and it is shown that desired link quality can be achieved by combining multiple streams pointing in slightly different directions from each other~\cite{suboptimaldiversity}. 
%\begin{figure*}[]
%\centering
%\includegraphics[width=6.0in,height=3.4in]{images/dualbandphotoshop.pdf}
%\caption{A schematic of dual band transmission.}
%\label{fig:dualbandphotoshop}
%\end{figure*}

Beamforming for the initial link setup, and beam-tracking to retain the desired quality communication link between moving devices are important for better user experience. A novel beam searching algorithm is proposed in~\cite{beamswitchingfast} which fasten the code book searching procedure specified in~\cite{beamcodebookharada}. Angle of arrival based approach is proposed to select the secondary beam if best beam is blocked due to obstacles in~\cite{AoAbeamswitching}. Xueli et al.~\cite{beamswitchingxeuli} have proposed a learning based beam switching algorithm if LOS path is blocked and proved that learning based approach is better than instantaneous decision based approach. Presently, literature on beam training and tracking is limited and lacks measurement based studies in deployments and thus require more efforts. 

%It is extremely important for moving device pairs to keep their beam aligned to achieve desired link quality.
IEEE 802.11ad also provides fallback option using fast session transfer to 2.4/5\,GHz channel. However, if multiple links fallback on 2.4/5\,GHz simultaneously, interference can limit the data transmission capabilities of each link. Specifically, if PCP/AP is involved in frequent switching from 60\,GHz to 2.4\,GHz, then other STAs which are able to communicate at 60\,GHz have to suffer unnecessarily. This is because a frame transmission at 2.4/5\,GHz would take about ten times more channel time than at 60\,GHz. 

Further, to realize the WLAN concept at 60\,GHz, multiple PCP/APs need to be installed to cover the indoor areas as different rooms separated by walls require their own PCP/AP. This makes network management a challenging task. In case of mobility, frequent association-disassociation events are triggered. Thus, device discovery or AP discovery becomes an important challenge due to directional communication. Therefore, management of multi-Gb/s 60\,GHz WLAN is an important challenge and requires novel approaches. Using 2.4/5\,GHz channel for transmission of control information -- with occasional fallback option for data transmission when 60\,GHz link is not available -- is an attractive alternative to provide seamless multi-Gb/s WLAN coverage. In such architectures, 60\,GHz can be used for data plane while 2.4\,GHz can be used for control plane. Device discovery, association and 60\,GHz channel reservation can be performed over 2.4\,GHz. This type of network architectures with opportunistic fallback of data plane communication to 2.4\,GHz could be a viable option for seamless WLAN experience. Mandke, et al.~\cite{multibandarch}, have discussed the motivation for a dual band WLAN operating simultaneously over 2.4\,GHz and 60\,GHz. Classification of traffic over the 2.4\,GHz and 60\,GHz frequency band is discussed. A 2.4\,GHz assisted 60\,GHz neighbour discovery and association mechanism are proposed in~\cite{multibanddiscovery}, and it is shown that with the help of 2.4\,GHz transmission, 60\,GHz device discovery procedure can be accelerated. However, Simultaneous operation over 2.4/5\,GHz frequency band (for control/management information transmission) and 60\,GHz frequency band (for data transmission) is not explored much and requires novel approaches for network architecture and MAC protocol design. Another important scheme for easier network management is the use of radio over fiber techniques to facilitate seamless communication at 60\,GHz\cite{RoFMAC}.  We see  need of further investigations that may lead to amendments to IEEE 802.11ad standard.
\subsection{Role of 60 GHz communication in 5G era}\label{role_5G}
Presently, researchers, academics and industries across the world are discussing the next generation mobile communication networks, i.e., 5G. One of the important focus of 5G is to introduce communication at new frequencies in the mmWave frequency bands (28-32, 38-42 and 55-67\,GHz) in order to deal with the scarcity of spectrum at lower frequency bands using large bandwidths available at mmWave bands~\cite{fiveG,Jefry5G}. Use of mmWave bands is aimed at indoor as well as outdoor environments.  In case of outdoors, 60\,GHz band is mainly suitable for wireless back-haul connectivity while 28-32\,GHz and 38-42\,GHz bands are suitable for mmWave cellular networks. However, recently, Zhu et al. have demonstrated that no such fundamental barriers exist which prevent the use of 60\,GHz in outdoor small cells~\cite{demystify60GHz}. It is shown that 60\,GHz can be easily used for outdoor mmWave cells of around 100\,m radius. So far, short range 60\,GHz propagation in indoor environment has been studied well in the last few years. Though 5G networks employing mmWave frequencies in outdoor environment have different propagation characteristics, many lessons can still be learnt from the already existing research on short range indoor communications at 60\,GHz. Further, integration of high speed mmWave communications with already existing cellular networks requires novel designs for physical interface for flexible spectrum access. Moreover, to enable seamless connectivity irrespective of environment, novel network architectures, integration of different protocols used in indoor or outdoor environments and offloading mechanisms between these protocols are required. One important difference with respect to existing system is that communication would highly rely on adaptive beamforming capabilities. Thus, link outage would be a major issue rather than interference. Also, Channel variation will be much faster because of extremely small channel coherence time at mmWave frequencies due to much higher Doppler spread. This will have a huge impact on cell search, broadcast signaling and multiple access schemes. One of the approach to solve these issues could be to allow the simultaneous connectivity with several base stations so that in case of mobility, connectivity with the base station can be maintained. 

To leverage the benefits of mmWave based outdoor and indoor networks, decoupling of control plane and data plane is seen as an important strategy which would lead to \textit{heterogeneity} in mmWave communications. Using LTE as the control and management plane for the closely spaced mmWave base stations could be a viable solution for network management, offloading and mobility management. All these require further research in terms of novel signaling and control procedures, MAC mechanisms, power saving techniques and transceiver design. 
\section{Conclusion}\label{conc}
In this article, we compared the three standards namely, ECMA-387, IEEE 802.15.3c and IEEE 802.11ad, which are proposed for 60\,GHz short range communication. We provided a brief account of these standards in terms of  medium access control mechanisms, beamforming procedures, device discovery  and network architectures. Standards were compared on the basis of different schemes proposed for robust WLAN/WPAN connectivity at 60\,GHz. IEEE 802,11ad has emerged as the choice for Gb/s WLAN because of its back compatibility with WiFi and several other provisions such as relay support and fallback options, etc. Further, we identified various still to be addressed issues for robust WLAN services at 60\,GHz frequency bands. In addition, the role of 60\,GHz communication in 5G networks was discussed. We believe that mmWave communications will become an integral part of 5G networks in indoor and outdoor environments. Its integration with the existing lower frequency networks requires considerable efforts at PHY and MAC layers. Already existing 60\,GHz research and standardizations can be helpful in defining new schemes for 5G networks.

\section*{Acknowledgment}
This work is funded by the Dutch national project SOWICI.
\bibliographystyle{IEEEtran}
\bibliography{referencessurvey60GHz}
\end{document}